\documentstyle[11pt]{article}
\newlength{\bredde}
\def\slash#1{\settowidth{\bredde}{$#1$}\ifmmode\,\raisebox{.15ex}{/}
\hspace*{-\bredde} #1\else$\,\raisebox{.15ex}{/}\hspace*{-\bredde} #1$\fi}
\newcommand{\beq}{\begin{equation}}
\newcommand{\eeq}{\end{equation}}
\newcommand{\ba}{\begin{array}{ccc}}
\newcommand{\ea}{\end{array}}

\newcommand{\noi}{\vspace{12pt}\noindent}
\newcommand{\lG}{\raise.3ex\hbox{$\stackrel{\leftarrow}{G}$}}
\newcommand{\lU}{\raise.3ex\hbox{$\stackrel{\leftarrow}{U}$}}
\newcommand{\lP}{\raise.3ex\hbox{$\stackrel{\leftarrow}{{\cal P}}$}}
\newcommand{\leta}{\raise.3ex\hbox{$\stackrel{\leftarrow}{\eta}$}}
\newcommand{\lOmega}{\raise.3ex\hbox{$\stackrel{\leftarrow}{\Omega}$}}
\newcommand{\ldr}{\raise.3ex\hbox{$\stackrel{\leftarrow}{\delta^r}$}}
\newcommand{\N}{{\mathcal{N}}}
\def\m2{{\mathcal{M}}^{\dagger}{\mathcal{M}}}
\def\mb2{M^2}

\def\beqn{\begin{eqnarray}}
\def\eeqn{\end{eqnarray}}

\def\gtwid{\raise.3ex\hbox{$>$\kern-.75em\lower1ex\hbox{$\sim$}}}
\def\ltwid{\raise.3ex\hbox{$<$\kern-.75em\lower1ex\hbox{$\sim$}}}

\begin{document}
\topmargin -1.4cm
\oddsidemargin -0.8cm
\evensidemargin -0.8cm
\title{\Large{{\bf Partially Quenched Chiral Condensates from the
Replica Method}}}

\vspace{1.5cm}

\author{~\\{\sc Poul H. Damgaard} \\~\\
The Niels Bohr Institute\\ Blegdamsvej 17\\ DK-2100 Copenhagen {\O}\\
Denmark}
\date{\today} 
\maketitle
\vfill
\begin{abstract} 
A large-$N_f$ expansion is used to compute the partially quenched
chiral condensate of QCD in the microscopic finite-volume scaling
region.
\end{abstract}
\vfill

\begin{flushleft}
NBI-HE-00-01 \\
hep-lat/0001002
\end{flushleft}
\thispagestyle{empty}
\newpage

\noi
While the replica method has found widespread use in condensed
matter physics as a means of generating quenched averages, it has
only rarely been applied in particle-physics contexts. Analytical
approaches to quenching -- so often used in lattice gauge theory
simulations -- have typically been based on the alternative
supersymmetric method \cite{BG} as applied to the effective chiral
Lagrangian (see also ref. \cite{S}). 
In the past few years this supersymmetric method has 
been successful
in deriving exact finite-volume scaling laws for the partially quenched
chiral condensate, and in computing analytically the smallest Dirac
operator spectrum in gauge theories with spontaneous chiral symmetry
breaking \cite{OTV}. Because the supersymmetric method relies on
a subtle definition of the global integration supermanifold of the
effective Lagrangian, it is nevertheless of interest if the same
results can be arrived at in an entirely different manner. Steps
in that direction were made very recently in ref. \cite{DS}, where it
was shown that small-mass and large-mass series expansions of the
partially quenched chiral condensate can be computed via 
the replica method. In the present paper we shall extend these results by
showing that also a large-$N_f$ expansion is a suitable starting point
for the replica method. The resulting series expansion for the partially
quenched chiral condensate turns out to be a resummed version of the
asymptotic large-mass expansion derived in ref. \cite{DS}, and thus more 
accurate in the appropriate regime. 

\noi
In the present context the replica method consists in adding to the
SU($N_c\geq 3$) QCD Lagrangian of 
$N_f$ physical quark flavors $N_v$ additional 
``valence quarks'', which, in the simplest case, are taken to be
mass-degenerate. We denote the physical quark masses by $m_f$, and
the mass of the valence quarks by $m_v$. We next consider the extended QCD
partition function in a sector of topological charge $\nu$, which we
for simplicity take to be non-negative (all physical results are in any
case symmetric under $\nu \to -\nu$):
\beq
{\cal Z}_{\nu}^{(N_{f}+N_{v})} ~=~ 
\left(\prod_{f=1}^{N_{f}} m_f^{\nu}\right)m_{v}^{N_{v}\nu}
\int\! [dA]_{\nu}
~{\det}'(i\slash{D} - m_v)^{N_{v}}\prod_{f=1}^{N_{f}}
{\det}'(i\slash{D} - m_f) ~e^{-S_{YM}[A]} ~. \label{Zoriginal}
\eeq
The prime on the determinants indicate that they exclude zero modes, whose
effect has already been taken into account by the prefactors.
This is an (unnormalized) average of $N_v$ identical replicas of the 
fermionic partition function
\beq
{\cal Z}_v ~\equiv~ \int\! d\bar{\psi}d\psi~ \exp\left[\int\! d^4 x 
\bar{\psi}(i\slash{D} - m_v)\psi\right] ~.
\eeq
Letting $N_v=0$ just reproduces the original QCD partition function. But
the theory extended with $N_v$ additional quark species can serve as
a generating functional of partially quenched averages of $\bar{\psi}\psi$
if one sets $N_v$ to zero {\em after} having performed the required
differentiations. Assuming that $\N \equiv N_f+N_v$ is small enough for 
spontaneous
chiral symmetry breaking to take place according to the conventional
SU($\N)\times$SU($\N) \to$ SU($\N$), the theory can be
analyzed in the low-energy region by means of its effective chiral
Lagrangian. For $N_v$ integer, this is completely straightforward, and
no additional assumptions are required. In order to apply the replica method
one must in addition rely on analyticity
in the number of flavors $\N$. This is the non-trivial step,
and it is a priori far from obvious that it can be done at all. But as
shown in ref. \cite{DS}, the needed analytic continuation is possible in
both small-mass and large-mass expansions. This is because 
the dimension of the coset group SU($\N$) in these expansions
appears only in simple
group invariants that are at most rational functions of polynomials in
$\N$. Except for the appearance of poles, such functions involving
only polynomials of $\N$ can unambiguously be continued to non-integer
$\N$. As in ref. \cite{DS}, we shall restrict ourselves to the
low-energy and finite-volume regime where $V \ll 1/m_{\pi}^4$ in the
limit where $V$ is sent to infinity. 
In this finite-volume scaling region there are also
exact non-perturbative results from both the supersymmetric method
\cite{OTV} and universal Random Matrix Theory formulas \cite{SV}
with which to compare our results. The conventional infinite-volume
chiral condensate is always normalized to $\Sigma$, independently of the 
physical number of flavors $N_f$.

\noi
The (mass-dependent) partially quenched chiral condensate in the given 
finite four-volume $V$ is defined by
\beq
\frac{\Sigma_{\nu}(\mu_v,\{\mu_f\})}{\Sigma} ~\equiv~ \lim_{N_{v}\to 0}
\frac{1}{N_{v}}\frac{\partial}{\partial \mu_{v}}
\ln {\cal Z}_{\nu}^{(\N)} ~. \label{sigmapq}
\eeq
As indicated, in the above scaling region the masses only enter in the
combinations $\mu_f\equiv m_fV\Sigma$ and
$\mu_v\equiv m_vV\Sigma$. In lattice gauge theory simulations the implied
finite-volume scaling
has been checked with staggered fermions \cite{VMC} and, very recently,
also with lattice fermions sensitive to gauge field topology \cite{DEHN}. 

\noi 
The effective finite-volume partition function is
\beq
{\cal Z}^{(\N)}_{\nu} = \int_{U(\N)}\! dU~ (\det U)^{\nu}
\exp\left[\frac{1}{2} {\rm Tr}({\cal M}
U^{\dagger} + U{\cal M}^{\dagger})\right]  ~, \label{ZUE}
\eeq
where ${\cal M}$ is the $\N\times\N$ matrix of 
rescaled masses:
$\mu_f = m_f\Sigma V$ for $f=1,\ldots, N_f$ and 
$\mu_v = m_v\Sigma V$ in the 
remaining $N_v$ entries. We shall here compute the partially quenched
chiral condensate by means of a large-$N_f$ expansion of this effective
partition function \cite{GN}. Because the effective action in eq. (\ref{ZUE})
involves $\N$ terms, 
while the integration is over $\N^2$
variables, it is conventional to perform the $1/\N$-expansion
by considering an action rescaled by $\N$:
\beq
{\cal Z}^{(\N)}_{\nu} = \int_{U(\N)}\! dU~ (\det U)^{\nu}
\exp\left[{\mathcal{N}}{\rm Tr}(A
U^{\dagger} + U{A}^{\dagger})\right]  ~. \label{ZUEGN}
\eeq
Letting $\lambda_a$ denote the eigenvalues of $A^{\dagger}A$, the relation
between these and the rescaled masses are thus
\beq
\lambda_a^{1/2} ~=~ \frac{1}{2\N}~\mu_a ~.
\eeq
At $N_f=\infty$, the starting point of
a large-$N_f$ expansion, the effective partition function is known
to describe two diffferent phases \cite{GN}. Defining
\beq
\sigma_k ~\equiv~ \frac{1}{\N}\sum_a \frac{1}{\lambda_a^{k/2}} ~,
\eeq
the transition between the two phases occurs at $\sigma_1 = 2$ for $\nu=0$
\cite{GN}, and this remains unchanged for any finite $\nu$.
We shall here focus on the phase with $\sigma_1 < 2$, which thus
corresponds to large masses. We also note that
\beq
\sigma_k ~=~ 2^k\N^{k-1}\sum_a \frac{1}{\mu_{a}^{k}} ~,
\eeq
and introduce in addition
\beq
\bar{\sigma}_k ~\equiv~ \sum_a \frac{1}{\mu_a^k} ~.\label{sigmabar}
\eeq
Conventionally one defines the free energy by $F^{(\nu)} 
\equiv (1/\N)\ln{\cal Z}_{\nu}$, but we shall instead need ${\cal F}^{(\nu)}
\equiv \N F^{(\nu)} = \ln{\cal Z}_{\nu}$ (see $e.g.$ eq. (\ref{sigmapq})).
In the large-$\N$ expansion one most readily computes 
\beq
F^{(\nu)}_a ~\equiv~ \frac{\partial F^{(\nu)}}{\partial\lambda_{a}}
~=~ \frac{2\N}{\mu_{a}}\frac{\partial{\cal F}^{(\nu)}}{\partial\mu_{a}} ~,
\label{Fderiv}
\eeq
which is particularly convenient for our purposes, since this is directly
related to the partially quenched chiral condensate:
\beq
\frac{\Sigma_{\nu}(\mu_v,\{\mu_f\})}{\Sigma} ~=~ \left.\frac{1}{N_v}
\frac{\partial}{\partial\mu_{v}}{\cal F}^{(\nu)}\right|_{N_{v}=0} ~.
\eeq
We thus need $N_v$-fold degenerate eigenvalues $\mu_v$, and can trivially
take the $N_v \to 0$ limit (as was to be expected in an expansion
around $N_f = \infty$).

\noi
In the sector of topological charge $\nu=0$ a large-$\N$ expansion of the
effective partition function (\ref{ZUEGN}) has
been worked out to high orders by Gross and Newman \cite{GN}, and we can 
directly make use of their results. They find, in the phase with
$\sigma_1 < 2$,
\beqn
F^{(0)}_a &=& \frac{1}{\lambda_a^{1/2}}\left(1 - \frac{1}{2\N}
\sum_b\frac{1}{\lambda_a^{1/2} + \lambda_b^{1/2}}\right) - \frac{1}{16\N^2
(2-\sigma_1)\lambda_a^{3/2}} \cr
&& - \frac{9}{256\N^4}\left[\frac{1}{(2-\sigma_1)^3\lambda_a^{5/2}}
+ \frac{\sigma_3}{(2-\sigma_1)^4\lambda_a^{3/2}}\right] \cr
&& - \frac{9}{2048\N^6}\left[\frac{25}{(2-\sigma_1)^5\lambda_a^{7/2}}
+ \frac{42\sigma_3}{(2-\sigma_1)^6\lambda_a^{5/2}}\right. \cr
&& + \left.\left(\frac{42\sigma_3^2}{(2-\sigma_1)^7} + 
\frac{25\sigma_5}{(2- \sigma_1)^6}\right)\frac{1}{\lambda_a^{3/2}}
\right] + \ldots
\eeqn
It is not a priory obvious that this large-$\N$ expansion is suitable
for our purposes, since we are certainly
only interested in the expansion for relatively small $N_f$. In fact,
we shall even also seek the other extreme limit of $N_f=0$, which
just corresponds to the fully quenched case. However, in the phase
with $\sigma_1 < 2$ the large-$\N$ expansion ought to be connected with the
saddle-point expansion, which in turn should be related to the large-mass
expansion derived in ref. \cite{DS}. As we shall explicitly verify below,
this is indeed the case. 

\noi
Using relations (\ref{sigmabar}) and (\ref{Fderiv}) we directly
recover an expansion for the partially quenched chiral condensate in a
gauge field sector of vanishing topological charge:
\beqn
\frac{\Sigma_{0}(\mu_v,\{\mu_f\})}{\Sigma} &=& 1 - \sum_{j=1}^{N_{f}}
\frac{1}{\mu_v+\mu_j} - \frac{1}{\mu_v^2}
\left[\frac{1}{8(1-\bar{\sigma}_1)}
+ \frac{9}{128(1-\bar{\sigma}_1)^4}\bar{\sigma}_3 \right.\cr
&& + \left.\frac{189}{512(1-\bar{\sigma}_1)^7}\bar{\sigma}_3^2 + 
\frac{225}{1024
(1-\bar{\sigma}_1)^6}\bar{\sigma}_5 + \ldots ~\right] \cr
&& - \frac{1}{\mu_v^4}\left[\frac{9}{128(1-\bar{\sigma}_1)^3} +
\frac{189}{512(1-\bar{\sigma}_1)^6}\bar{\sigma}_3 + \ldots ~\right] \cr
&& - \frac{1}{\mu_v^6}\left[\frac{225}{1024(1-\bar{\sigma}_1)^5} + \ldots
 ~\right] + \ldots \label{slargenf}
\eeqn
Remarkably, all explicit factors of $\N$ have cancelled, and the above
expansion is perfectly well-defined for any $N_f$, even all the way down
to $N_f=0.$\footnote{Note that the limit $N_v=0$ has already been taken here,
so that $\N = N_f$ at this point.} The only requirement
is that $\sigma_1 < 2$, which translates into $\bar{\sigma}_1 < 1$ (in the
$N_f=0$ case there is no such requirement, and one simply sets all
void sums to zero in the above expansion). As
expected, the resulting series is thus related to the large-mass
expansion of ref. \cite{DS}. In fact, many curious regularities observed
in the large-mass expansion of ref. \cite{DS} now find their explanation: the
above large-$\N$ series is a non-trivial {\em resummation} of the
ordinary large-mass expansion that highlights the special role played
by the first inverse mass sum $\bar{\sigma}_1$. In ref. \cite{DS} the
large-mass series was for convenience truncated at an order (7th)
taken to be the same both in the quenched mass $\mu_v$ and the physical
masses $\mu_f$. Indeed, if one truncates the expansion of eq. (\ref{slargenf})
at the same order, one recovers exactly the large-mass series derived
in ref. \cite{DS}. It is also in this form that one can most easily
compare with the analytical results of both the supersymmetric method
\cite{OTV} and Random Matrix Theory \cite{VMC}, and as already noted
in ref. \cite{DS} there is perfect agreement with both. However, only
the large-$N_f$ expansion has revealed that a resummation of the series
is possible.

\noi
The large-mass expansion of ref. \cite{DS} made use of the fact that
the $\nu=0$ effective partition function factors into two pieces, one
originating directly from the leading-order saddle point solution, 
and a remaining function
which is annihilated by a set of Virasoro generators. Unfortunately,
it is not presently
known how to extend this technique based on Virasoro constraints 
to the case of $\nu \neq 0$. But the 
large-$\N$ expansion is not afflicted with this problem, and it
can easily be carried through to the case $\nu \neq 0$. 
Also here much of the work has already been done for us.
In particular, Brihaye and Rossi \cite{BR} have derived the large-$\N$
expansion up to order $1/\N^4$ in the free energy. Using our notation 
their result for the free energy $F^{(\nu)}=1/\N \ln{\cal Z}_{\nu}$ 
reads
\beqn
F^{(\nu)} &=& 2\sum_a \lambda_a^{1/2} -\frac{1}{2\N}\sum_{a,b}
\ln[\lambda_a^{1/2}+\lambda_b^{1/2}] \cr
&& + \frac{1}{2\N}(\nu^2-\frac{1}{4})\ln(2-\sigma_1)
+ \frac{(\nu^2-\frac{1}{4})(\nu^2-\frac{9}{4})}{24\N^3}
\frac{\sigma_3}{(2-\sigma_1)^3} + \ldots
\eeqn
This corresponds to 
\beqn
F^{(\nu)}_a &=& \frac{1}{\lambda_a^{1/2}}\left(1 - \frac{1}{2\N}
\sum_b\frac{1}{\lambda_a^{1/2} + \lambda_b^{1/2}}\right) - \frac{1}{16\N^2
(2-\sigma_1)\lambda_a^{3/2}}
+ \frac{\nu^2-\frac{1}{4}}{4\N^2(2-\sigma_1)\lambda_a^{3/2}} \cr
&& -\frac{(\nu^2-\frac{1}{4})(\nu^2-\frac{9}{4})}{16\N^4}
\left[\frac{\sigma_3}{(2-\sigma_1)^4\lambda_a^{3/2}} +
\frac{1}{(2-\sigma_1)^3\lambda_a^{5/2}}\right] + \ldots ~,
\label{Fanu}
\eeqn
which in turn, introducing the rescaled sums (\ref{sigmabar}) and the 
identification (\ref{Fderiv}),
leads to a partially quenched chiral condensate computed up
to ${\cal O}(1/\N^6)$:
\beqn
\frac{\Sigma_{\nu}(\mu_v,\{\mu_f\})}{\Sigma} &=& 1 - \sum_{j=1}^{N_{f}}
\frac{1}{\mu_v+\mu_j} + \frac{1}{\mu_v^2}
\left[\frac{4\nu^2-1}{8(1-\bar{\sigma}_1)}
- \frac{(4\nu^2-1)(4\nu^2-9)}{128(1-\bar{\sigma}_1)^4}\bar{\sigma}_3 +\ldots
 ~\right] \cr
&& - \frac{1}{\mu_v^4}\left[\frac{(4\nu^2-1)(4\nu^2-9)}
{128(1-\bar{\sigma}_1)^3} + \ldots ~\right] + \ldots \label{slargenfnu}
\eeqn
Of course, for $\nu=0$ we recover the leading orders of the expansion
(\ref{slargenf}). We also note that the first terms in eq. (\ref{Fanu}), 
those from the 
classical saddle point, are unaffacted by the presence of 
$(\det U)^{\nu}$ in the integrand, as expected. If we expand the result 
(\ref{slargenfnu}) in $\bar{\sigma}_1$ we find
\beqn
\frac{\Sigma_{\nu}(\mu_v,\{\mu_f\})}{\Sigma} &=& 1 - \sum_{j=1}^{N_{f}}
\frac{1}{\mu_v+\mu_j} + \frac{1}{\mu_v^2}
\left[\frac{4\nu^2-1}{8}\left(1 + \bar{\sigma}_1 + \bar{\sigma}_1^2 + 
\bar{\sigma}_1^3 + \bar{\sigma}_1^4 + \ldots \right)\right.\cr
&&- \left.\frac{(4\nu^2-1)(4\nu^2-9)}{128}\bar{\sigma}_3\left(1 + 
4\bar{\sigma}_1 
+ 10\bar{\sigma}_1^2 + 20\bar{\sigma}_1^3 + \ldots\right)~  +\ldots ~\right]
\cr
&& - \frac{1}{\mu_v^4}\frac{(4\nu^2-1)(4\nu^2-9)}{128}\left[1 + 
3\bar{\sigma}_1 + 6\bar{\sigma}_1^2 + 10\bar{\sigma}_1^3 + \ldots ~\right]  
+\ldots  \label{slargenfnuexp}
\eeqn
As a check, this agrees with the asymptotic expansion
of the the fully quenched ($N_f=0$) chiral condensate \cite{VMC,OTV}
\beq
\frac{\Sigma_{\nu}(\mu_v)}{\Sigma} ~=~ \mu_v\left[I_{\nu}(\mu_v)
K_{\nu}(\mu_v) + I_{\nu+1}(\mu_v)K_{\nu-1}(\mu_v)\right] + 
\frac{\nu}{\mu_{v}} ~, \label{sigmaNf0}
\eeq
as obtained from either the supersymmetric method or from Random Matrix
Theory. However, the standard asymptotic
expansions of these expressions, using the individual asymptotic 
series of the
modified Bessel function $I_n(x)$ and $K_n(x)$, miss the possibility
of resummations as in the large-$\N$ expansions 
(\ref{slargenf}) and (\ref{slargenfnu}). It is interesting to note that
the otherwise ubiquitous $\nu/\mu_v$-term in the partially quenched chiral
condensate, which normally \cite{D} can be traced directly back to the 
$\mu_v^{N_{v}\nu}$ factor from the zero modes in the original QCD 
partition function (\ref{Zoriginal}),
is missing in the above expansion. It is due to a precise cancellation
between this term and a ${\cal O}(1/\mu_v)$ piece arising from the
asymptotic expansions of the modified Bessel functions in (\ref{sigmaNf0}),
which thus, beyond the classical saddle-point terms, starts 
at ${\cal O}(1/\mu_v^2)$.\footnote{This cancellation was first observed by 
J. Verbaarschot (private communication).}   

\noi
To conclude, the large-$\N$ expansion of the effective
partition function of QCD in the finite-volume scaling regime $V \ll
1/m_{\pi}^4$ is a good starting point for an analytical computation
of the partially quenched chiral condensate by means of the replica method.
The resulting series is a resummation of the large-mass expansion derived
by means of the same replica method in ref. \cite{DS}. In the $\nu=0$ sector
we have explicitly checked this agreement, while for sectors with $\nu \neq 0$
we have confirmed in the $N_f=0$ case that the resulting series agrees with 
the asymptotic
expansion of the partially quenched chiral condensate as it obtained
by both the supersymmetric technique and Random Matrix Theory. The replica
method is clearly a suitable alternative to these methods, at least
in series expansions. To go beyond such expansions 
requires an analytical handle on ``external-field'' group integrals for
unitary groups U($n$) with $n$ extended beyond integers, a problem that
may appear forbiddingly difficult. Fortunately, less is actually needed:
only the exact coefficient of the leading term in an $\epsilon$-expansion
of the group U($N_f + \epsilon$), where $N_f$ is integer. Finding this
exact expression will allow one to establish the surprising relationship
between such group integrals and those of the supergroup extensions
discussed in ref. \cite{OTV}. Other outstanding questions are how to use
this method to derive spectral correlation functions of the Dirac operator
\cite{OTV}, and how to understand why the correlations of the
smallest Dirac eigenvalues can be described in terms of a theory extended with 
additional fermion species \cite{AD} (a result that has a natural 
explanation in the supergroup formulation \cite{OTV}). Finally,
the replica method may also find
use as an alternative way to perform partially quenched chiral perturbation
theory in general, $i.e$, beyond the finite-volume scaling regime considered
here. 

\noi
{\sc Acknowledgement:}\\
This work was supported in part by EU TMR grant no. ERBFMRXCT97-0122.

\end{document}